\documentclass[conference]{IEEEtran}
\usepackage{cite}
\usepackage{amsmath,amssymb,amsfonts}
\usepackage{algorithmic}
\usepackage{graphicx}
\usepackage{textcomp}
\usepackage{xcolor}
\usepackage{hyperref}
\def\BibTeX{{\rm B\kern-.05em{\sc i\kern-.025em b}\kern-.08em
    T\kern-.1667em\lower.7ex\hbox{E}\kern-.125emX}}
\begin{document}

\title{Knowledge Retrieval With Functional Object-Oriented Networks}

\author{\IEEEauthorblockN{Shawn Diaz}}

\maketitle

\begin{abstract}
In this experiment, three different search algorithms are implemented for the purpose of extracting a task tree from a large knowledge graph, known as the Functional Object-Oriented Network (FOON). Using a universal FOON, which contains knowledge extracted by annotating online cooking videos, and a desired goal, a task tree can be retrieved. The process of searching the universal FOON for task tree retrieval is tested using iterative deepening search and greedy best-first search with two different heuristic functions. The performance of these three algorithms is analyzed and compared. The results of the experiment show that iterative deepening performs strongly overall. However, different heuristics in an informed search proved to be beneficial for certain situations.

\end{abstract}

\section{Introduction}
The development of search algorithms for the Functional Object-Oriented Network (FOON) is an important component in the knowledge retrieval process of extracting a subgraph, called a task tree; This task tree can be used by robots to complete manipulation tasks. The FOON is a structured knowledge representation presented by Paulius et al. which is capable of generating a sequence of motions for a robot to follow in order to carry out desired tasks \cite{4}. A task tree can utilize knowledge from several sources to produce a new task sequence which has not been seen before \cite{5}. When extracting a task tree, a search is done which looks at a list of items present in the space (i.e. a kitchen). From here, the search should be able to find functional units that will lead to inputs already in the kitchen. When the task tree saves these functional units from the search, the complete task tree is reversed. This gives a structure that a robot can use to create a desired item from start to finish. 

The overall performance of the search algorithms used, as well as the structures used for knowledge representation, is important to be mindful of since the FOON can possibly be generalized for use with other objects. When dealing with tasks where the objects involved are familiar, the performance of the knowledge retrieval process is not as critical. However, if the goal is to generalize the FOON beyond the kitchen and apply it to other objects, performance becomes more important. Paulius et al. explore this more in depth by looking at situations where the universal FOON does not contain information about an object in a specific state \cite{3}. This leads to a situation where the task tree retrieval process is halted and cannot achieve the desired goal. Instead, the idea would be to use the limited knowledge contained in a FOON to transfer knowledge from one item to another. After exploring two different means of generalization, they concluded that there would need to be developments in how the robot interacts with items by specifying how objects should be used to solve a given task.

\section{Video Annotation and FOON Creation}
In order for a FOON to be created, annotations of cooking videos must be made first. The annotation process makes note of objects, and their states before and after some motion within a scene. Each motion or action is contained within a structure called a functional unit, which also contains the information about the object states before and after an action. When one or more functional units are placed together in sequence, a subgraph is formed, which represents an activity. From here, a universal FOON can be created, which contains many sequences of completed activities from the video annotation process. 

Once a universal FOON is created through the annotation process, a knowledge retrieval process can take place. This process allows a robot to retrieve a subgraph from the universal FOON; This subgraph is called a task tree. The knowledge retrieval process for generating a task tree accounts for a list of items in its environment, such as a kitchen. The manner in which it searches for a solution is dependent on the algorithm used (e.g. depth-first search, breadth-first search, iterative deepening, etc.). Certain algorithms may be better performing than others in certain contexts. This is discussed further in \autoref{sec:method}.

\begin{figure}[htbp]
    \includegraphics[width=\columnwidth]{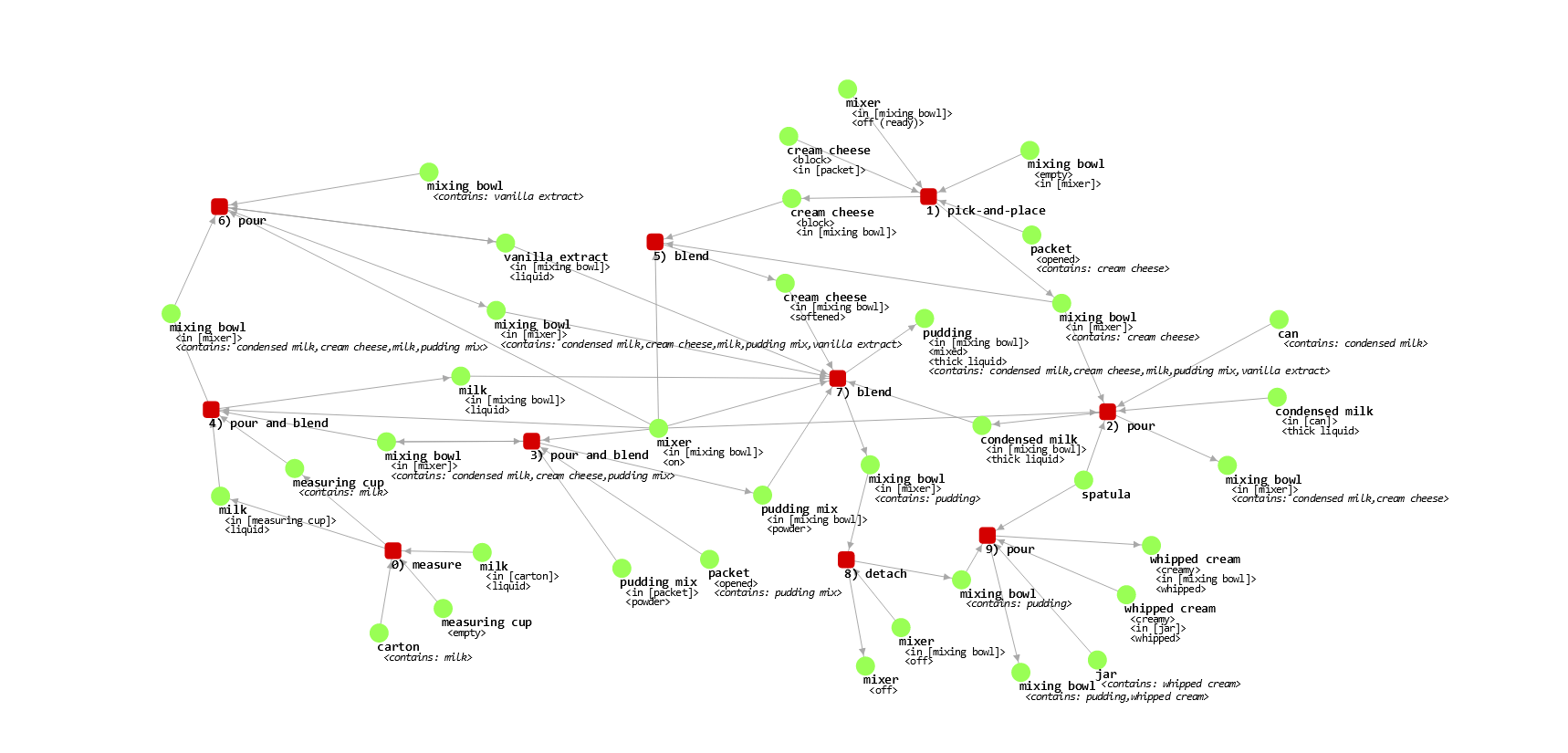}
    \caption{An example of a FOON subgraph showing the steps necessary to make whipped cream.}
    \label{fig}
\end{figure}

\section{Method}
\label{sec:method}
Three algorithms were implemented and analyzed in the experiment. The first algorithm was an uninformed search algorithm called iterative deepening search. With iterative deepening, a graph is defined, along with a variable depth and a goal node. First, the goal node is found in the universal FOON. Then, a check is done to see if there are input nodes which are not found in the kitchen (the kitchen is a JSON file with all of the items and their states in the kitchen). Then, the inputs nodes not found in the kitchen are searched recursively with depth-first search until a solution is found (i.e. the leaf nodes are available in the kitchen). In the case of iterative deepening, if there are multiple candidate functional units, the first unit is always selected. This process runs up to the current iteration's defined depth, before the depth is incremented and depth-first search is called again with the new maximum depth. This is repeated until the goal is reached (i.e. the leaf nodes are available in the kitchen).

The second and third algorithms that were implemented were greedy best-first search algorithms, each with a different heuristic function. Greedy best-first search is an informed search algorithm (i.e. based on a heuristic function) which extends paths based greedily on a given heuristic. Similar to the iterative deepening search algorithm, the goal node is found in the universal FOON. Once the goal node is found, the input nodes of its functional unit are placed into a queue if they are not found in the kitchen. From here, the candidate units for each node are found and the best unit based on the heuristic function is picked. The first heuristic is defined as the success rate of the motion from each candidate unit. These rates are given in a text file, which was previously parsed. Some of these motions can be seen in \autoref{tab2}. For instance, if we have two candidate units: mix and stir, the function will search the motions text file for each motion node and compare their success rates. Whichever motion has a higher success rate will be selected; In this case, it would select mix. More information on this heuristic is given in related works by Paulius et al. \cite{2}.

\begin{table}[htbp]
\caption{Example of Motions and Their Success Rates}
    \begin{center}
        \begin{tabular}{|c|c|}
            \hline
            \textbf{Motion Label} & \textbf{Success Rate}\\
            \hline
            Chop & 0.10\\
            \hline
            Pour & 0.90 \\
            \hline
            Mix & 0.90  \\
            \hline
            Crack & 0.20 \\
            \hline
            Pick-and-place & 0.80 \\
            \hline
            Stir & 0.80 \\
            \hline
            Bake & 0.40 \\
            \hline
            \end{tabular}
    \label{tab2}
    \end{center}
\end{table}

The third algorithm, which is also a greedy best-first search algorithm, uses a heuristic function which counts the number of input objects in the candidate functional units, then selects the one with the lowest count. It works very similarly to the first heuristic algorithm, in that it greedily selects functional units based on the heuristic function. For example, if scrambled eggs can be made with egg, oil, cheese, and onion or just egg, oil, and salt, the latter option would be selected. By counting the number of input objects, and selecting the candidate unit with the least amount, the idea would be that the search would reach the desired goal sooner since there are less nodes to process overall.

\section{Experiment and Discussion}
\subsection{Algorithm Effectiveness}
After running each of the search algorithms and generating task trees for several recipes, an analysis can be done on the performance of each algorithm to determine their overall effectiveness in the knowledge retrieval process. Looking at the comparison between the number of functional units generated in the task tree for each recipe in \autoref{tab1}, the informed search algorithms (i.e. heuristic 1 and heuristic 2) tend to do better in some situations where the recipe is more complex, and therefore may contain more steps to complete. With simpler recipes (e.g. Ice or Sweet Potato), the number of functional units is lower, rendering the informed search algorithm less useful. When using an uninformed search algorithm such as iterative deepening search, rather than selecting candidate functional units based on a heuristic function, the first candidate is always selected. This leads to some instances where iterative deepening produces a task tree with less functional units than the heuristic-based search algorithms, and some cases where a task tree is produced with more functional units. In an ideal situation, all candidate functional units would be searched to find the optimal solution. To keep it simple, this was not done.

\begin{table}[htbp]
\caption{Algorithms Comparison (Functional Unit Count)}
    \begin{center}
        \begin{tabular}{|c|c|c|c|}
            \hline
            \textbf{Goal}&\multicolumn{3}{|c|}{\textbf{No. of Functional Units}}\\
            \cline{2-4} 
            \textbf{Node} & \textbf{Iterative Deepening} & \textbf{Heuristic 1} & \textbf{Heuristic 2}\\
            \hline
            Scrambled Egg & 29 & 40 & 32\\
            \hline
            Hashbrown & 32 & 41 & 26\\
            \hline
            Greek Salad & 28 & 32 & 32\\
            \hline
            Whipped Cream & 10 & 15 & 15\\
            \hline
            Macaroni & 7 & 7 & 7\\
            \hline
            Sweet Potato & 3 & 3 & 3\\
            \hline
            Ice & 1 & 1 & 1\\
            \hline
            \end{tabular}
    \label{tab1}
    \end{center}
\end{table}

\subsection{Complexity Analysis}
A FOON is created with an adjacency list, making the retrieval time for candidate functional units $\mathcal{O}(1)$. The worst case is that the number of candidates is the total number of recipes in universal FOON. For $n$ functional units in the task tree, the time required is $\mathcal{O}(n)$. When generating the task tree, if no equivalent ingredient exists in the task tree already, then a search would have to be done through other recipes in FOON to find an equivalent ingredient. This would typically take $\mathcal{O}(k)$ time, where $k$ is the number of ingredients in FOON. However, since a mapping was created from pre-processing, finding an equivalent ingredient is $\mathcal{O}(1)$, and overall complexity is $\mathcal{O}(n)$ \cite{1}. To keep things simple, in the case of iterative deepening search, the selected candidate function was always the first one in the list. And with the informed search algorithms, the heuristics deal with information already present (i.e. motion success-rate and input node count). These algorithms would not be as effected with increases in universal FOON, since they would not necessarily explore all potential paths. The greedy best-first search algorithms will greedily select the best candidate unit based on the heuristic, and the iterative deepening search algorithm will select the first candidate unit every time.

\bibliographystyle{IEEEbib}
\bibliography{refs}

\begin{thebibliography}{1}

\bibitem{4}
David Paulius, Yongqiang Huang, Roger Milton, William~D. Buchanan, Jeanine Sam,
  and Yu~Sun,
\newblock ``Functional object-oriented network for manipulation learning,''
\newblock in {\em 2016 IEEE/RSJ International Conference on Intelligent Robots
  and Systems (IROS)}, 2016, pp. 2655--2662.

\bibitem{5}
Md~Sadman Sakib, Hailey Baez, David Paulius, and Yu~Sun,
\newblock ``Evaluating recipes generated from functional object-oriented
  network,'' 2021.

\bibitem{3}
David Paulius, Ahmad~B. Jelodar, and Yu~Sun,
\newblock ``Functional object-oriented network: Construction \& expansion,''
\newblock in {\em 2018 IEEE International Conference on Robotics and Automation
  (ICRA)}, 2018, pp. 5935--5941.

\bibitem{2}
David Paulius, Kelvin Sheng~Pei Dong, and Yu~Sun,
\newblock ``Task planning with a weighted functional object-oriented network,''
\newblock in {\em 2021 IEEE International Conference on Robotics and Automation
  (ICRA)}, 2021, pp. 3904--3910.

\bibitem{1}
Md.~Sadman Sakib, David Paulius, and Yu~Sun,
\newblock ``Approximate task tree retrieval in a knowledge network for robotic
  cooking,''
\newblock {\em IEEE Robotics and Automation Letters}, vol. 7, no. 4, pp.
  11492--11499, 2022.

\end{thebibliography}

\end{document}